\documentclass[11pt,reqno]{amsart}
\usepackage{amsmath,amsgen,amsbsy,amstext,amsopn,amsthm,amssymb}
\newtheorem{thm}{THEOREM}

\newtheorem{lem}{LEMMA}[section]

\theoremstyle{definition}

\newtheorem{rem}{REMARK}
\newcommand{\infspec}{{\rm inf\ spec\ }}
\newcommand{\R}{{\mathbb R}}

\newcommand{\C}{{\mathbb C}}

\newcommand{\A}{{\mathcal A}}
\newcommand{\Da}{{D}^\ast}

\newcommand{\F}{{\mathcal F}}
\newcommand{\Q}{{\mathcal Q}}
\newcommand{\Ll}{L}

\newcommand{\Hh}{{\mathcal H}}

\newcommand{\aan}{a_{\lambda}}
\newcommand{\ac}{a^{\ast}_{\lambda}}

\newcommand{\ean}{\varepsilon^{\lambda}}
\newcommand{\half}{\mbox{$\frac{1}{2}$}}

\newcommand{\al}{{\alpha}}

\newcommand{\as}{\sqrt{\alpha}}

\newcommand{\Ow}{{\mathcal O}}
\newcommand{\ora}{{|0\rangle}}
\newcommand{\lao}{{\langle 0|}}

\newcommand{\la}{\Lambda}

\newcommand{\Hf}{H_{\mathrm f}}
\newcommand{\tHf}{\tilde H_{\mathrm f}}

\def\qed{\hfill$\diamondsuit$}
\numberwithin{equation}{section}

\begin{document}

\title[Enhanced binding]
{Enhanced binding revisited for a  spinless particle  in   non-relativistic QED}

\thanks{C.H.
    acknowledges support through the European Union's IHP
    network Analysis \& Quantum HPRN-CT-2002-00277.}

\author[I. Catto, P. Exner \& Ch. Hainzl]{Isabelle Catto, Pavel Exner and Christian Hainzl}
\address{CEREMADE, CNRS UMR 7534, Universit\'e Paris-Dauphine, Place
du Mar\'e\-chal de Lattre de Tassigny, F-75775 Paris Cedex 16,
France} \email{catto@ceremade.dauphine.fr}
\address{Department of Theoretical Physics, NPI, Academy of Sciences,
25068 \v{R}e\v{z}--Prague, Czechia}
\email{exner@ujf.cas.cz}
\address{CEREMADE, Universit\'e
  Paris-Dauphine, Place du Mar\'echal de Lattre de Tassigny, F-75775 Paris \& Laboratoire de Math\'ematiques
Paris-Sud-Bat 425, F-91405 Orsay Cedex} \email{hainzl@ceremade.dauphine.fr}

\date{\today}
%%%%%%%%%%%%%%%%%%%%%%%%%
\begin{abstract}
We consider a spinless particle coupled to a quantized Bose field
and show that such a system has a ground state for two classes of
short-range potentials which are alone too weak to have a
zero-energy resonance.
\end{abstract}
%%%%%%%%%%%%%%%%%%%%%%%%%

\maketitle

\section{INTRODUCTION AND MAIN RESULTS}

\noindent Let us begin by specifying the classes of potentials to
be considered. Throughout the paper we suppose that (i) $\:V\in
L^\infty$ is nonzero and attractive, $V\le 0$, and (ii) $\:V$
belongs to $L^{3/2}(\R^3)$, which is well known to ensure, in
particular, that $V$ lies in the Rollnik class, \emph{i.e.}
  % ------------- %
 \begin{equation*}
 \Vert V\Vert_R^{\,2}:=\iint_{\R^6}\frac{|V(x)|\,|V(y)|}
 {|x-y|^2}\,dxdy<+\infty\;.
 \end{equation*}
 % ------------- %
Finally, we adopt one of the following assumptions: (iii) $\:V$ is
\emph{strictly} attractive, $V<0$, and satisfies the inequality
 % ------------- %
\begin{equation}\label{eq:hypV}
|\Delta\,V|\leq C\,|V|\;,
\end{equation}
 % ------------- %
with a positive constant $C$, or alternatively (iii') $\:V$ is
compactly supported with $\Delta V$ integrable. We will say more
on these assumptions in the remarks following Theorem~\ref{mth}
below.

We denote by $H_{\lambda}$ the family of Schr\"odinger operators
$H_{\lambda}:= p^2+\lambda\,V$ on $\Ll^2(\R^3)$ for positive
parameters $\lambda$. Its eigenvalues are monotonically decreasing
functions of $\lambda$ in $[0,+\infty) $, and it is well-known
that at some positive critical value $\lambda_{0}$, which is
called the ``coupling-constant threshold" an eigenvalue emerges
from the continuum. More precisely, if $\lambda\leq\lambda_{0}$,
$H_{\lambda}$ has no negative eigenvalues, whereas when
$\lambda>\lambda_{0}$ it has at least one bound state. \vskip6pt
We now couple this Schr\"odinger operator to the radiation field
and  consider the so-called \emph{Pauli-Fierz operator}
 % ------------- %
\begin{equation}\label{rpf}
H_{\alpha}^V = (p + \sqrt{\al}\,A(x))^2  + H_{\rm f}+\lambda\, V\;,
\end{equation}
 % ------------- %
acting on  the Hilbert space
 % ------------- %
\begin{equation*}
\Hh = \Ll^{2}({\mathbb {R}}^{3};\C)\otimes \F
\end{equation*}
 % ------------- %
where $\F=\bigoplus_{n=0}^{+\infty}\Ll_b^2(\R^{3n};\C)$ is the
Fock space for the photon field  and $\Ll_b^2(\R^{3n})$ is the
space of symmetric functions in $\Ll^2(\R^{3n})$ representing
$n$-photon states, with $n=0$ corresponding to the one-dimensional
vacuum sector. Following the usual convention we abuse notation
and use the same symbol for $H_{\rm f}$ and $\overline{I\otimes
H_{\rm f}}$, etc. The operator $H_{\alpha}^V$ is essentially
self-adjoint on $\mathcal{D}(\Delta)\cap \mathcal{D}(H_{\rm f})$,
where the symbol $\mathcal{D}$ denotes the operator domain --
see~\cite{BFS}.

\vskip6pt We denote the ground state energy as
 % ------------- %
\begin{equation}\label{def:gs}
E(\alpha,\lambda\,V):=\infspec H_{\alpha}^V\;,
\end{equation}
 % ------------- %
and the spectrum is then the half-line
$[E(\alpha,\lambda\,V),\,+\infty)$, see again~\cite{BFS}. In
\cite[Theorem 3.1]{GLL}, Griesemer, Lieb and Loss show under
rather weak assumptions about the potential $V$ that, in the case
when $\lambda>\lambda_{0}$, \emph{i.e.} when the Sch\"odinger
operator $-\Delta+\lambda \,V$ has a ground state, it persists
after coupling to the radiation field. Moreover, in~\cite{HVV},
Hainzl, Vougalter and Vugalter prove for a particular class of
potentials that in the case when $\lambda=\lambda_{0}$, the
coupling to the field can create a ground state for small coupling
constant $\alpha$, despite the fact that the underlying
Sch\"odinger operator $-\Delta+\lambda_{0} \,V$ has no ground
state. Recall also that in the case of a particle with spin the
same result was proved by two of the present authors in~\cite{CH};
a different proof is given in \cite{CVV}.

\vskip6pt Our goal here is to show that $E(\alpha,\lambda\,V)$ is
an eigenvalue of $H_{\alpha}^V $ for coupling in some interval
$(\gamma,\lambda_{0}]$, with $\lambda_0 - \gamma =\Ow(\al)$ for
$\alpha$ small. The strategy of proof will be based on the
expansion of the self-energy in powers of $\alpha$, as provided in
\cite{H1,H2,HVV,CH}, and by checking that the Griesemer-Lieb-Loss
criterium (\textit{i.e.}, inequality~\eqref{eq:crit} below)
 is satisfied for $\alpha$ small enough.
To this aim we use a result of Klaus and Simon \cite{KS}, which
allows us to demonstrate the enhanced binding for the class of
potentials indicated above; we need not assume like in \cite{HVV} that the
Hamiltonian has a zero-energy resonance in the absence of the
field.

\vskip6pt Recall that in the dipole approximation, in the case of
a large coupling $\al$ the enhanced binding was shown earlier by
Hiroshima and Spohn \cite{HiSp}, see also \cite{AA} in the
context of linear coupling.

%%%%%%%%%%%%
\vskip6pt We fix units in such a way that the Planck constant
$\hbar=1$, the speed of light $c=1$, and the electron mass
$m=\half$. The electron charge is then given by $e=\as$, with
$\alpha\approx 1/137$ being the fine structure constant. In the
present paper $\al$ plays the role of a small, dimensionless
number which measures the coupling to the radiation field. Our
results hold for {\em sufficiently small values \/} of $\al$. The
operator $p= -\mathrm{i}\nabla$ is the electron momentum while $A$
is the quantized magnetic vector potential, which  is given by
 % ------------- %
\begin{equation*}
A(x) = \sum_{\lambda = 1,2} \int_{\R^3}
\frac{\chi(|k|)}{2\pi\,|k|^{1/2}} \,\ean(k)\big[ \aan(k)
\mathrm{e}^{\mathrm{i}k\cdot x}
+\ac(k) \mathrm{e}^{-\mathrm{i}k\cdot x}\big] \mathrm{d}k\;,
\end{equation*}
 % ------------- %
where the annihilation and creation operators $\aan$ and  $\ac$,
respectively,  satisfy the usual commutation relations
 % ------------- %
\begin{equation*}
[a_\nu (k),\ac(q)] = \delta(k-q)\delta_{\lambda,\nu}\;,
\end{equation*}
 % ------------- %
and
 % ------------- %
\begin{equation*}
[\aan(k), a_\nu(q)] = 0, \ \ [\ac(k),a_\nu^\ast(q)] = 0\;.
\end{equation*}
 % ------------- %
In the following we use the notation
 % ------------- %
\begin{equation}
A(x)= D(x) + D^*(x).
\end{equation}
 % ------------- %
The vectors $\ean(k) \in \R^3$ in $A(x)$ are orthonormal
polarization vectors perpendicular to $k$ which are chosen in a
such a way that
 % ------------- %
\begin{equation}\label{convention}
     \varepsilon^2(k)=\frac{k}{| k|}\wedge
     \varepsilon^1(k)\;.
\end{equation}
 % ------------- %
The function $\chi(|k|)$ describes the ultraviolet cutoff for the
interaction at large wave-numbers $k$. For the sake of simplicity
we choose for $\chi$ the Heaviside function $\Theta(\Lambda
-|k|)$; more general cut-off functions would work, however, let us
emphasize the fact that we shall sometimes use the radial symmetry
of  $\chi$ in the proofs. Throughout the paper we assume $\la=1$.
This corresponds to the energy $\,m\,c^2$ in our system of units
and represents a natural upper bound to which the validity of
non-relativistic QED can be extended.

The photon field energy $H_{\mathrm{f}}$ is given by
 % ------------- %
\begin{equation}
H_{\mathrm{f}} = \sum_{\lambda= 1,2} \int_{\R^3}
|k| \ac (k) \aan (k) \mathrm{d}k
\end{equation}
 % ------------- %
and the field momentum reads
 % ------------- %
\begin{equation}
P_{\mathrm{f}} = \sum_{\lambda= 1,2} \int_{\R^3}
k \ac (k) \aan (k) \mathrm{d}k\;.
\end{equation}
 % ------------- %
Since $\Hh$ can be also written as $\bigoplus_{n=0}^{+\infty}
\Ll^{2}({\mathbb {R}}^{3};\C)\otimes \Ll_b^2(\R^{3n};\C)$ we can
express a general vector  $\Psi \in \Hh$ as a direct sum
 % ------------- %
\begin{equation}
\Psi= \bigoplus_{n=0}^\infty\: \psi_n\;,
\end{equation}
 % ------------- %
where $\psi_n = \psi_n(x,k_1,\dots,k_n)$ is an $n$-photon state.
For simplicity, we do not include the variables corresponding to
the polarization of the photons.

To simplify further the notation, we observe that the Hamiltonian
is translation invariant and introduce the unitary transformation
 % ------------- %
\begin{equation}\label{defU}
U = \mathrm{e}^{\mathrm{i}P_{\mathrm{f}} \cdot x}
\end{equation}
 % ------------- %
acting on $\Hh$. Since
 % ------------- %
\[U
A(x) U^*= A(0)
\]
 % ------------- %
and
 % ------------- %
\begin{equation}
UpU^*=p-P_{\mathrm{f}}
\end{equation}
 % ------------- %
we obtain
 % ------------- %
\begin{equation}\label{defTU}
U\,H_{\alpha}^V\,U^* = \big(p-P_{\mathrm{f}} + \as A\big)^2
+ H_{\mathrm{f}}+\lambda\,V\;,
\end{equation}
 % ------------- %
where $A=A(0)$. The operator $U$ preserves spectral properties, in
particular,
 % ------------- %
\begin{equation}
\infspec\big[U\,H_{\alpha}^V\,U^*] = \infspec H_{\alpha}^V\;.
\end{equation}
 % ------------- %
Thus we shall rather work with $U\,H_{\alpha}^V\,U^*$ in the
following; abusing the notation we will use again the symbol
$\,H_{\alpha}^V\,$ for it.

\vskip10pt Our main result is the following
%%%%%%%%%%%%%%%%%%%
\begin{thm} \label{mth}
Adopt the assumptions (i), (ii), and either (iii) or (iii'). Then
there exists a function $g:\: \R^+\to\; (0,1)$ such that for any
small enough $\alpha$ and all $\lambda\in\; (\lambda_{0}\,
\big(1\!-\!g(\alpha)\big), \lambda_{0}]$, the spectral threshold
$E(\alpha ,\lambda\,V)$ is an eigenvalue of $H_{\alpha}^V$.
 \end{thm}

%%%%%%%%%%%%%%%%%%%%%
\vskip10pt
\begin{rem}
The assumptions combine different type of requirements. For
instance, (\ref{eq:hypV}) or $\Delta V\in L^1$ impose restrictions
mainly on local regularity of the potential. On the other hand,
(ii) regulates its decay; on a heuristic level one may say that
the potential should behave as $|x|^{-2-\epsilon}$ at infinity.
\end{rem}

%%%%%%%%%%%%%%%%%%%%%
\begin{rem}
It comes out of the proof that $g(\alpha)$ is of order of
$\alpha$, more specifically, the relation (\ref{eq:defg}) shows
that $g(\alpha)= c\alpha + \mathcal{O}(\alpha^2 \ln\al)$ with
$c>0$ holds as $\alpha\to 0$. It is important that we get in this
way an asymptotical \emph{lower} bound to $g(\alpha)$ which allows
us to assess how much the binding is enhanced.
\end{rem}

%%%%%%%%%%%%%%%%%%%%%
\begin{rem}
In connection with the previous remark we want to emphasize that
all the constants appearing in the proof can be evaluated
explicitly. Assuming that we choose a potential $V$ such that the
constants $C, b(V)$ in \eqref{eq:hypV} and \eqref{aV},
respectively, are of order of one, it turns out that
Theorem~\ref{mth} holds for $\alpha \lesssim 10^{-2}$ which covers
the physically important case.
\end{rem}

%%%%%%%%%%%%%%%%%%%%%
\begin{rem}We have recalled above the results \cite{GLL} and
\cite{HVV}, the latter using the existence of a zero-energy
resonance state together with a continuity argument which shows
that $H_{\alpha}^V$ has a bound state for values of $\lambda$
slightly below $\lambda_{0}$. Our strategy is similar but the
proof is more constructive, in particular, it provides a rough
estimate on how far below $\lambda_{0}$ one can descend to still
ensure the existence of a ground state. In addition our method
covers a different and in several respects wider class of
potentials $V$, in particular, we require neither a compact
support nor the radial symmetry of the potential.
\end{rem}

%%%%%%%%%%%%%%%%%
\begin{rem}
Using the methods of \cite{CH}, Theorem \ref{mth} can also be
proven for the case of particles with spin. Unfortunately the
numbers of inequalities needed increase dramatically. For that
reason we restricted our attention to the more convenient case of
bosons.
\end{rem}

%%%%%%%%%%%%%%%%%%%%%%
\section{PROOF OF THEOREM 1}
%%%%%%%%%%%%%%%%%%%%%%

\noindent Let $0<\lambda\leq \lambda_{0}$. According to Griesemer,
Lieb and Loss \cite{GLL} the ground state exists provided
 % ------------- %
\begin{equation}\label{eq:crit}
E(\alpha ,\lambda\,V)<E(\alpha, 0)\,,
\end{equation}
 % ------------- %
where $E(\alpha, 0)$ is the electron self-energy. Thus we are
going to construct a trial state $\Psi\in \Ll^2(\R^3)\otimes\F$
which ensures that the last inequality is satisfied,
 % ------------- %
\begin{equation}\label{eq:GLL}
(\Psi; H_{\alpha}^V\,\Psi)<E(\alpha,0)\,\Vert\Psi\Vert^2\,.
\end{equation}
 % ------------- %
The strategy of proof is as follows: we will compare the
respective expansions of $E(\alpha, 0)$ and $(\Psi;
H_{\alpha}^V\,\Psi)$ for the trial state $\Psi$ in terms of the
coupling constant $\alpha$. From \cite{H1,HVV} we already know the
first three terms in the Taylor expansion of the former, namely
 % ------------- %
\begin{equation}\label{eq:self}
\left\vert E(\alpha,0)-\alpha\,\pi^{-1}+\alpha^2\,\lao D\, D
\A_{\alpha}^{-1}\Da\,\Da\ora\right\vert\leq C_{\rm self}\,\alpha^3
\;,\end{equation}
 % ------------- %
for some positive constant $C_{\rm self}$, where $\ora$ is the
vacuum vector, $\langle\cdot;\cdot\rangle$ denotes the scalar
product in the photon Fock space $\F$, and with
  % ------------- %
\begin{equation}\label{eq:Aal}
\A_{\alpha}= P_{\rm f}^2+H_{\rm f}+2\,\alpha\,\Da\, D\;.
\end{equation}
 % ------------- %
Recall that actually in \cite{HVV} the proof is given for $\A_0$
instead of $\A_{\alpha}$ in the second-order term
in~\eqref{eq:self}, but the same argument carries through
\textsl{mutatis mutandis} to the present case.

Consider now a quantity $g(\alpha)\in (0,1)$, to be determined
later, and observe that when the coupling parameter $\lambda $
satisfies
 % ------------- %
 $$ \big(1\!-\!g(\alpha)\big)\,\lambda_{0}< \lambda\leq \lambda_{0},$$
 % ------------- %
then the Schr\"odinger operator
 % ------------- %
 $$h_\alpha^\lambda:=-\big(1\!-\!g(\alpha)\big)\,p^2+\lambda\,V$$
 % ------------- %
has a negative eigenvalue $e_{\lambda}(\alpha):= -|e_{\lambda}|$
at the bottom of the spectrum. This trivially follows from the
inequality $\frac{\lambda} {1-g(\alpha)}>\lambda_{0}$ and our
choice of $\lambda_0$ to be critical. We denote by
$\psi_{\lambda}$ a corresponding eigenstate which may be chosen
without loss of generality as real-valued and normalized in
$\Ll^2$. Our trial function, to be inserted into \eqref{eq:GLL},
will involve only two photons being of the form
 % ------------- %
\begin{subequations}
\begin{equation}\label{trial}
\Psi=\psi_{\lambda}\oplus\psi_{1}\oplus\psi_{2}
\end{equation}
 % ------------- %
with
 % ------------- %
\begin{equation}\label{trial1}
\psi_1=-2\,\as\,L^{-1}\Da\,p\,\psi_{\lambda}
\end{equation}
 % ------------- %
and
 % ------------- %
\begin{equation}\label{trial2}
\psi_2=-\,\al\,L^{-1}\Da\,\Da\,\psi_{\lambda}\;.
\end{equation}
\end{subequations}
 % ------------- %
The operator $L$ on $\Ll^2(\R^3)\otimes\mathcal{F}$ appearing in
here is defined by
 % ------------- %
\begin{equation}\label{eq:L}
L=\big(1\!-\!g(\alpha)\big)\,(p-P_{\rm
f})^2+\lambda\,V+|e_{\lambda}|+H_{\rm f}+2\,\alpha\,\Da\, D\;;
\end{equation}
 % ------------- %
the definitions (\ref{trial1}) and (\ref{trial2}) make sense
because $L$ is invertible on the orthogonal complement of the
vacuum sector $\Ll^2(\R^3)\otimes\mathbb{C}\ora$; this follows
from the fact that it is unitarily equivalent to $h_\alpha^\lambda
+|e_{\lambda}|+H_{\rm f}+2\,\alpha\,\Da(x) D(x)$ by means of the
operator (\ref{defU}). Note that, with the abuse of notation
mentioned in the opening, we often use $\psi_\lambda$ as a shorthand
for $\psi_\lambda\otimes\ora$.

The canonical commutation relations yield the identity
 % ------------- %
$$ A^2=\Da\,\Da+D\,D+2\,\Da\, D+\frac{1}\pi\;;$$
 % ------------- %
using it together with commutativity of $P_\mathrm{f}$ with $D,\,
D^*$, we find that
 % ------------- %
\begin{multline}
\label{sack} (\Psi;H_{\alpha}^V\Psi) =g(\alpha)\,\Vert
P\Psi\Vert^2+ \left\lbrack\frac{\alpha}\pi
-|e_\lambda|\right\rbrack\,\Vert\Psi\Vert^2 \\
+ (\Psi;L\Psi) + 2\Re (\Psi;[2\as PD^*+\al D^*D^*]\Psi)
+(\psi_\lambda; \big(h_\alpha^\lambda+|e_\lambda|\big)\,
\psi_\lambda)\,,
\end{multline}
 % ------------- %
where $P:= p-P_\mathrm{f}$ denotes the total momentum; the last
term in the right-hand side of \eqref{sack} cancels by definition
of $e_\lambda$, $h_\alpha^\lambda$ and $\psi_\lambda$.

Let us further remark that the argument will require -- similarly
as in the previous works \cite{HS,CH,HHS} -- to replace the field
Hamiltonian $\Hf$ by $\tHf:=\Hf + \alpha^3$ in order to deal with
the logarithmically divergent infrared terms; this would amount to
adding an extra $-\alpha^3\, \Vert\Psi\Vert^2$ term at the
right-hand side of \eqref{sack}. Observe that by our choice of
$\Psi$ the identity
 % ------------- %
\begin{multline}
(\Psi;L\Psi) +2\Re (\Psi;[2\as PD^*+\al D^*D^*]\Psi) =
(\psi_\lambda;L \psi_\lambda) \\
-2 \Re (\psi_\lambda;-\alpha DD L^{-1}(-2\as PD)^* L^{-1}
(-2\as PD)^* \psi_\lambda)
\\ - \|L^{-1/2}[2\as PD^*+\al D^*D^*]\psi_\lambda\|^2
\end{multline}
 % ------------- %
holds, and the same is true if we replace $L$ in the last relation
by $\tilde{L}$ referring to $\tHf$. Thus we obtain
 % ------------- %
\begin{multline}\label{energy}
(\Psi; \tilde H_{\alpha}^V\,\Psi)= g(\alpha)\,\Vert P\Psi\Vert^2 +
\left[ \frac{\alpha}\pi - |e_\lambda| -\al^3\right]\|\Psi\|^2
-4\,\al\,\Vert \tilde{L}^{-1/2}\Da\,p\, \psi_\lambda\Vert^2\\
-\al^2\,(\psi_\lambda;D\,D \tilde{L}^{-1}\Da\,\Da\psi_\lambda)
+\,8\,\al^2\,\Re(\tilde{L}^{-1}\Da\,\Da \psi_\lambda;P\, \Da
\tilde{L}^{-1}\Da\,p\psi_\lambda)\,,
\end{multline}
 % ------------- %
where $\tilde H_{\alpha}^V$ refers again to $\tHf$. On the other
hand, apart of taming the infrared singularity the extra term is
irrelevant, as long as we looking for an effect of order of
$\alpha^2$. This is why we will abuse notation writing non-tilded
quantities everywhere except the one place below where this
Hamiltonian shift indeed matters.

To estimate the last term at the right-hand side of
\eqref{energy}, notice that the Cauchy-Schwarz inequality yields
 % ------------- %
\begin{multline}\label{term1}
8\,\al^2\,\big\vert(L^{-1}\Da\,\Da \psi_\lambda;P\,
\Da L^{-1}\Da\,p\psi_\lambda)\big\vert \\
\leq\frac{4\,\al^3}{a}\, \Vert L^{-1/2}P\,DL^{-1}\Da\,\Da
\psi_\lambda\Vert^2 +4\,\al\,a\,\Vert
L^{-1/2}\Da\,p\,\psi_\lambda\Vert^2
\end{multline}
 % ------------- %
with a positive constant $a$ to be chosen later. The last term can
be combined with the similar term in \eqref{energy} giving
 % ------------- %
\begin{multline}\label{energybis}
(\Psi; H_{\alpha}^V\,\Psi)\leq g(\alpha)\,\Vert P\Psi\Vert^2
-|e_\lambda|\,\Vert\Psi\Vert^2+\frac{\alpha}\pi\,\Vert\Psi\Vert^2
-4\,\al\,(1-a)\,\Vert L^{-1/2}\Da\,p\, \psi_\lambda\Vert^2\\
-\al^2\,(\psi_\lambda;D\,D L^{-1}\Da\,\Da\psi_\lambda)+
\frac{4\,\al^3}{a}\,\Vert L^{-1/2}P\,DL^{-1}\Da\,\Da
\psi_\lambda\Vert^2\, .
\end{multline}
 % ------------- %
To estimate further the terms appearing in \eqref{term1} we need a
series of technical lemmata.

\begin{lem}\label{lem:1} The following inequality holds,
 % ------------- %
\begin{equation}\label{fait1}
(\psi_\lambda;D\,D L^{-1}\Da\,\Da\psi_\lambda) \geq \lao D\,D
\A_{\alpha}^{-1} \Da\, \Da\ora+g(\al)\,C_1\,,
\end{equation}
 % ------------- %
where
 % ------------- %
\[
C_{1}:= \Vert P_{\rm f}\A_{\alpha}^{-1}\Da\,\Da\ora\Vert^2\,.\]
 % ------------- %
\end{lem}
%%%%%%%%%%%%
\vskip6pt\noindent
\proof We denote $L=\Q+b$, with
\[
\Q= \big(1\!-\!g(\alpha)\big)\,(p^2+P_{\rm
f}^2)+\lambda\,V+|e_{\lambda}|+H_{\rm f}+2\,\alpha\,\Da\, D
\]
and $b=-2\,\big(1\!-\!g(\alpha)\big)\,p P_{f} $, and we use twice
the second resolvent equation,
 % ------------- %
\begin{equation}\label{resolv}
(\Q+b)^{-1}=  \Q^{-1} -\Q^{-1}\,b\,\Q^{-1}
+\Q^{-1}\,b\,(\Q+b)^{-1}\,b\,\Q^{-1} \geq
\Q^{-1}-\Q^{-1}\,b\,\Q^{-1}\,,
\end{equation}
 % ------------- %
where the inverse of $\Q+b$ and the last inequality makes sense in
the complement to the vacuum sector where the operator is strictly
positive as we have remarked above. Hence we have
 % ------------- %
\begin{multline}
(\psi_\lambda;D\,D L^{-1}\Da\,\Da\psi_\lambda)
\geq(\psi_\lambda;D\,D \Q^{-1}\Da\,\Da\psi_\lambda)
\\+2\,\big(1\!-\!g(\alpha)\big)\,(\psi_\lambda;D\,D \Q^{-1}p P_{\rm f}
 \Q^{-1}\Da\,\Da\psi_\lambda)\;.
\end{multline}
 % ------------- %
Furthermore, the second term at the right-hand side vanishes. To
check this claim, recall that $\psi_\lambda$ belongs by
construction to the null-space of
$h_{\alpha}^\lambda+|e_{\lambda}|$, and that $h_{\alpha}^\lambda$
commutes with the operator
 % ------------- %
\[
\mathcal{K} :=  \big(1\!-\!g(\alpha)\big)\,P_{\rm f}^2+H_{\rm
f}+2\,\alpha\,\Da\, D.
\]
 % ------------- %
It follows easily that
 % ------------- %
\begin{equation}\label{eq:K}
\Q^{-1}\Da\,\Da\psi_\lambda=\mathcal{K}^{-1}\Da\,\Da\psi_\lambda\;,
\end{equation}
 % ------------- %
and therefore
 % ------------- %
\[
(\psi_\lambda;D\,D \Q^{-1}p P_{\rm f}
 \Q^{-1}\Da\,\Da\psi_\lambda)=
(\psi_\lambda;p\,\psi_\lambda) \lao D\,D \mathcal{K}^{-1}
 P_{\rm f}\mathcal{K}^{-1}\Da\,\Da\ora=0\;,\]
 % ------------- %
because $\psi_\lambda$ is real-valued as indicated above making
the first factor zero.

Using \eqref{eq:K} again, we find that the first term reads
 % ------------- %
\begin{multline}
(\psi_\lambda;D\,D \Q^{-1}\Da\,\Da\psi_\lambda)
=(\psi_\lambda;D\,D \mathcal{K}^{-1}\Da\,\Da\psi_\lambda)\\
=\Vert\psi_\lambda \Vert^2\,\lao D\,D \mathcal{K}^{-1}\Da\,\Da\ora\\
\geq\lao D\,D \mathcal{A}_{\alpha}^{-1}\Da\,\Da\ora
+g(\alpha)\,\lao D\,D \mathcal{A}_{\alpha}^{-1}P_{\rm f}^2
\mathcal{A}_{\alpha}^{-1}\Da\,\Da\ora\;,
\end{multline}
 % ------------- %
where in the last line we used the fact that $\psi_\lambda$ is
normalized together with the second resolvent equation and
positivity of $\mathcal{K}$ in the complement of the vacuum
sector. Hence the claim. \qed

%%%%%%%%%%%%%%%%%%%%%%%%%%%%%%%%
\begin{lem} For any positive constant $\mu$ we have
 % ------------- %
\begin{eqnarray}\label{eq:mu}
\lefteqn{\Vert L^{-1/2}\Da\,p\, \psi_\lambda\Vert^2}\\
&\geq &[C_4(\mu)+\mu\,C_3(\mu)]\,\Vert p\,\psi_\lambda\Vert^
2-\frac{\lambda}{2}\,C_3(\mu)\,(\psi_\lambda;\Delta V\,\psi_\lambda)
\;,\nonumber
\end{eqnarray}
 % ------------- %
with
 % ------------- %
\begin{equation}\label{eq:C3}
C_3(\mu)=\frac 23 \lao D\big(\mathcal{K}\!+\!\mu\big)^{-2}\Da\ora
\end{equation}
 % ------------- %
and
\begin{equation}\label{eq:C4}
C_4(\mu)=\frac 23 \lao
D\big(\mathcal{K}\!+\!\mu\big)^{-1}\Da\ora\;.
\end{equation}
 % ------------- %
\end{lem}
%%%%%%%%%%%%%%%
\proof Using the relation analogous to \eqref{resolv}, we find
 % ------------- %
\begin{eqnarray*}
\lefteqn{p\,D L^{-1}\Da\,p\geq p\,D
\big(\mathcal{K}\!+\!\mu\big)^{-1}\Da\,p}\\
&&-p\,D \big(\mathcal{K}\!+\!\mu\big)^{-1} \Big[ -2\,p P_{\rm
f}\,\big(1\!-\!g(\alpha)\big)+h^\lambda_{\alpha}
+|e_\lambda|-\mu\Big]\big(\mathcal{K}\!+\!\mu\big)^{-1}\Da\,p\;.
\end{eqnarray*}
 % ------------- %
Again, since $\psi_{\lambda}$ is real valued the term containing
$p P_{\rm f}$ vanishes; notice that the same conclusion can also
be made using symmetry of the cut-off function. Since
$\mathcal{K}$ acts on photon variables whereas
$h^\lambda_{\alpha}$ acts on those of the electron, the two
operators commute and the second term at the right-hand side of
the last estimate can be rewritten as
 % ------------- %
\begin{multline}
(\psi_{\lambda};p\,D \big(\mathcal{K}\!+\!\mu\big)^{-1} \Big[
-2\,p P_{\rm
f}\,\big(1\!-\!g(\alpha)\big)+h^\lambda_{\alpha}+|e_\lambda|-
\mu\Big]\big(\mathcal{K}\!+\!\mu\big)^{-1}\Da\,p \psi_{\lambda})\\
=C_{3}(\mu)\,\big(p\psi_{\lambda};(h^\lambda_{\alpha}+|e_\lambda|-\mu)
\, p\psi_{\lambda}\big)
\end{multline}
 % ------------- %
with $C_{3}(\mu)$ given by \eqref{eq:C3}. Observing that
$\psi_{\lambda}$ belongs to the null-space of
$h^\lambda_{\alpha}+|e_\lambda|$ we can further cast a part of the
last expression into the form
 % ------------- %
\begin{equation}
(p\psi_{\lambda};(h^\lambda_{\alpha}+|e_\lambda|)  \,
p\psi_{\lambda})= -\frac 12 (\psi_\lambda;
[p,[p,h^\lambda_{\alpha}+|e_\lambda|]]\psi_\lambda)= \frac \lambda
2 (\psi_\lambda;\Delta V \psi_\lambda)
\end{equation}
 % ------------- %
to obtain
 % ------------- %
\[
\big(p\psi_{\lambda};(h^\lambda_{\alpha}+|e_\lambda|-\mu)  \,
p\psi_{\lambda}\big)=-\mu\,\Vert
p\psi_{\lambda}\Vert^2+\frac\lambda{2}\,\big(\psi_{\lambda};\Delta
V\,\psi_{\lambda}\big)\;.\]
 % ------------- %
On the other hand,
 % ------------- %
\[
(\psi_{\lambda};p\,D \big(\mathcal{K}\!+\!\mu\big)^{-1}\Da\,p
\psi_{\lambda})=C_{4}(\mu)\,\Vert p\psi_{\lambda}\Vert^2
\]
 % ------------- %
with $C_{4}(\mu)$ given by \eqref{eq:C4}, which concludes the
proof. \qed
 \vskip10pt
%%%%%%%%%%%%%%%%%%
Our next auxiliary result is the following:
 % ------------- %
\begin{lem}\label{lem:LT} Let $\alpha>0$ and
$0<\lambda\leq\lambda_{0}$, then for any positive number
$g(\alpha)<1\!-\frac{\lambda}{\lambda_{0}}$ there are positive
constants $\beta$ and $C(V)$ such that
 % ------------- %
\begin{equation}\label{eq:interpol}
p^2\leq \beta\,(h_\al^\lambda+|e_\lambda|)+C(V)\;.
\end{equation}
 % ------------- %
\end{lem}
 % ------------- %
\proof For the inequality \eqref{eq:interpol} to hold, the
constant $\beta$ which appears at the right-hand side has
obviously to satisfy the inequality $\beta>\frac{1}{1-g(\alpha)}$,
or equivalently $\beta(1\!-\!g(\alpha))-1>0$. We fix an arbitrary
$\beta$ with this property, to be specified later. Next we notice
that inequality \eqref{eq:interpol} will follow from
 % ------------- %
\begin{equation*}
p^2+\lambda\,\tilde V\geq -\tilde C(V)
\end{equation*}
 % ------------- %
with
 % ------------- %
\begin{equation*}
\tilde V:= \frac{1}{1\!-\!g(\alpha)-\beta^{-1}}\,V\;, \qquad
\tilde C(V):=\frac{C(V)}{\beta\,(1\!-\!g(\alpha))-1}\;,
\end{equation*}
 % ------------- %
because the last inequality is equivalent to \eqref{eq:interpol}
with the term $|e_\lambda|$ at the right-hand side neglected. In
other words, it is sufficient that the Schr\"{o}dinger operator
$p^2+\lambda\,\tilde V$ has no spectrum below $-\tilde C(V)$. From
the proof of the Birman-Schwinger bound in~\cite[Theorem
XIII.10]{bi:RS} and the fact that $\tilde V$ is non-positive it
follows that this happens if and only if
 % ------------- %
 \begin{equation}\label{eq:BS}
 \frac {\lambda^2 }{16\,\pi^2}\,\iint_{\R^6}\frac{|\tilde V(x)|\,
 |\tilde V(y)|}{|x-y|^2}\, \mathrm{e}^{-2\,\sqrt{\tilde C(V)}\,|x-y|}
 \,dxdy<1\;.
 \end{equation}
 % ------------- %
Let us denote by $\mathcal{K}_m$ the function $x\mapsto
\frac{\mathrm{e}^{-\sqrt{m}\, |x| }}{4\,\pi\,|x| ^2}$ with a fixed
positive $m$ which represents the resolvent kernel in the above
expression; it is clear that $\mathcal{K}_m$ belongs to
$L^1(\R^3)$ and $\displaystyle{\int_{\R^3}}
\mathcal{K}_m(x)\,dx=\frac{1}{\sqrt m}$. We employ these
observations in the following chain of inequalities,
 % ------------- %
\begin{eqnarray*}%\label{eq:BS2}
 \lefteqn{\frac {\lambda^2 }{16\,\pi^2}\,
 \iint_{\R^6}\frac{|\tilde V(x)|\,|\tilde V(y)|}{|x-y|^2}\,
 \mathrm{e}^{-2\,\sqrt{\tilde C(V)}\,|x-y|^2}\,dxdy}\\
 &= & \frac {\lambda^2 }{4\,\pi\,\big(1\!-\!g(\alpha)-\beta^{-1}\big)^2}\,
 \int_{\R^3}\big( V\star K_{4\,\tilde C(V)}\big)(x)\, V(x)\,dx\\
  &\leq & \frac {\lambda^2 }{4\,\pi\,\big(1\!-\!g(\alpha)-\beta^{-1}\big)^2}\,
  \Vert V\Vert_{L^2}^2\,\Vert K_{4\,\tilde C(V)}\Vert_{L^1}\\
  &\leq & \frac {\lambda^2 }{8\,\pi\,\big(1\!-\!g(\alpha)-\beta^{-1}\big)^2\,
  \sqrt{\tilde C(V)}}\,\Vert  V\Vert_{L^2}^2\;,
 \end{eqnarray*}
  % ------------- %
where in the second and third line we used Cauchy-Schwarz and
Young inequalities, respectively. Thus the bound \eqref{eq:BS}
will be satisfied if the last expression is smaller than one. For
a fixed $\beta> 2(1\!-\!g(\alpha))^{-1}$ we can estimate
$\big(1\!-\!g(\alpha) -\beta^{-1}\big)^{-1/2}$ by $\sqrt{\beta}$,
and consequently, the inequality \eqref{eq:interpol} will hold
uniformly in $\lambda\in [0;\lambda_0)$, as long as the positive
constant $C(V)$ is chosen large enough to satisfy
 % ------------- %
\begin{equation}\label{eq:choice}
\frac {\lambda_0^2\;\sqrt{\beta}\;\Vert V\Vert_{L^2}^2\,
}{8\,\pi\,\big(1\!-\!g(\alpha)-\beta^{-1}\big)^{3/2}}<\sqrt{\tilde
C(V)}\;,
\end{equation}
 % ------------- %
what we set out to prove. \qed \vskip10pt
%%%%%%%%%%%%%%%%%%%%%%%%%%%%%%%%

Note that the constants can be chosen explicitly. The left-hand
side of \eqref{eq:choice} diverges as $\beta\to\infty$ and in the
allowed interval it has a unique minimum at $\beta=\frac{4}{1\!-\!
g(\alpha)}$ where it attains the value $\frac
{2\,\lambda_0^2\;\Vert V\Vert_{L^2}^2\,}
{3\,\sqrt{3}\,\pi\,\big(1\!-\!g(\alpha)\big)^{2}}$. In other
words, the lemma is valid for this $\beta$ and
 % ------------- %
 $$
 C(V) =\left(\frac{2\,\lambda_0^2\;\Vert V\Vert^2_{L^2}\,}
 {3\pi\,(1\!-\!g(\alpha))^2} \right)^2\,.
 $$
 % ------------- %
The reader may wonder that we have not used here fully our
assumptions about the potential because for a bounded function
$V\in L^2$ is a weaker requirement than $V\in L^{3/2}$, however,
without the latter our main premise about existence of the
coupling constant threshold may not be valid.

\vskip10pt
%%%%%%%%%%%%%%%%%%%%%%%%%%%%%%%%

 % ------------- %
\begin{lem}\label{lem:faits34}
The following estimates hold:
 % ------------- %
\begin{equation}\label{fait3}
\Vert P\,L^{-1}\Da\,\Da\psi_\lambda\Vert^2\leq C_5
\end{equation}
 % ------------- %
and
 % ------------- %
\begin{equation}\label{fait4}
\Vert P\,L^{-1}\Da\,p\,\psi_\lambda\Vert^2\leq C_6(\al) \Vert
p\,\psi_\lambda\Vert^2 \end{equation}
 % ------------- %
with positive $C_5$ and $C_6(\al)$ given in \eqref{C5} and
\eqref{C6} below, depending on $\beta$ and $C(V)$ of the previous
lemma. Using the shifted Hamiltonian $\tHf:=\Hf + \alpha^3$, we
have $C_6(\al) \sim \ln(\al^{-3})$ as $\al\to 0+\:$.
\end{lem}
 % ------------- %
\proof By means of \eqref{eq:interpol} we get the estimates
 % ------------- %
\begin{multline}
L^{-1}P^2L^{-1} \leq L^{-1}\big[\beta(h_\al^\lambda+|e_\lambda|) +
C(V)\big]L^{-1}  \leq \frac{\beta}{2}\,H_{\rm f}^{-1} + C(V)
\Hf^{-2},
\end{multline}
 % ------------- %
valid in the appropriate part of the state space, namely when
sandwiched between vectors annulated by $P_\mathrm{f}$; in the
second inequality we used the fact that for any pair of commuting
operators $B,C$ with $C$ strictly positive we have
$(B+C)^{-1}B(B+C)^{-1} \le \frac{1}{2}C^{-1}$. In this way we
arrive at
 % ------------- %
\begin{equation}\label{C5}
\begin{split}
\Vert P\,L^{-1}\Da\,\Da\psi_\lambda\Vert^2 & \leq \Big[\frac
\beta2 \langle 0| DD \Hf^{-1}D^*D^*|0\rangle
+ C(V)\langle 0| DD \Hf^{-2}D^*D^*|0\rangle \Big]\\
& :=  C_5
\end{split}
\end{equation}
 % ------------- %
and
 % ------------- %
\begin{equation}\label{C6}
\begin{split}
\Vert P\,L^{-1}\Da\,p\psi_\lambda\Vert^2 & \leq  \|p
\psi_\lambda\|^2  \frac 23\Big[\frac \beta2 \langle 0| D
\Hf^{-1}D^*|0\rangle
+ C(V)\langle 0| D \Hf^{-2}D^*|0\rangle \Big]\\
& := \|p\psi_\lambda\|^2 C_6(\al)\,.
\end{split}
\end{equation}
 % ------------- %
Now we come to the place where the shift matters because without
it the right-hand side of \eqref{C6} is infrared divergent. With
the replacement $\Hf \to \Hf+\al^3$ we have
 % ------------- %
\begin{eqnarray}\nonumber
\lefteqn{\langle 0| D [\Hf+\al^3]^{-2}D^*|0\rangle }\\
&&= 8\pi \int_{0}^1d|k|\: \frac{|k|}{[|k|+\al^3]^2}= 8\pi
\big(\ln(\al^{-3}) + \al^3 - 1\big)\,,
\end{eqnarray}
 % ------------- %
which concludes the argument. \qed \vskip10pt
%%%%%%%%%%%%%%%%%%%%%%%%%%%%%%%%

Finally we come to our last technical result.
 % ------------- %
\begin{lem}\label{lem:fait99}
Under the assumption (iii') there is a positive constant $C$ such that
 % ------------- %
\begin{equation}\label{altern}
\left| (\psi_\lambda; \Delta V \psi_\lambda)\right| \le C\,
(\psi_\lambda; |V| \psi_\lambda)\,.
\end{equation}
 % ------------- %
\end{lem}
 % ------------- %
\proof Since $V\in L^\infty$ the ground state is represented on
$G:=\mathrm{supp\,}V$ by a positive smooth function. Hence
$\rho(\lambda):= \sup_G \psi_\lambda \, (\inf_G
\psi_\lambda)^{-1}$ makes sense and satisfies
$1\le\rho(\lambda)<\infty$; the same is true for $\lambda=
(1-g(\alpha))\lambda_0$ corresponding to the zero-energy
resonance. Using the standard estimate \cite[Thm.~IX.28]{bi:RS}
one can check that $\lambda\mapsto \psi_\lambda$ is continuous in
the $\|\cdot\|_\infty$ norm which implies continuity of the
function $\rho$. Consequently, there are positive $m$ and $M$ such
that
 % ------------- %
\begin{equation}\label{twoside}
0< m \le \tilde\psi_\lambda(x) \le M <\infty
\end{equation}
 % ------------- %
holds for all $x\in G$, $\:\lambda \in\; ]\lambda_{0}\,
\big(1-g(\alpha)\big); \lambda_{0}]$, and a suitable family of
non-normalized solutions (for $\psi_\lambda$ both the infimum and
supremum vanish, of course, as we approach the zero-energy
resonance). It follows that
 % ------------- %
\begin{equation}\label{upper}
\big(\tilde\psi_\lambda; \Delta V \tilde\psi_\lambda\big) \le M^2
\|\Delta V\|_{L^1}\,,
\end{equation}
 % ------------- %
while $(\tilde\psi_\lambda; |V| \tilde\psi_\lambda)\ge m^2
\|V\|_{L^1}$ is positive, so it can majorize \eqref{upper} when
multiplied by a sufficiently large $C$. \qed \vskip10pt

%%%%%%%%%%%%%%%%%%%%%%%%%%%%%%%%%%

Now we are ready to complete the proof of the theorem. From the
definitions of $\psi_{1}$ and  $\psi_{2}$ and with the help of
\eqref{fait3} and \eqref{fait4} we get
 % ------------- %
\begin{equation}\label{evalP}
g(\alpha)\,\Vert P\Psi\Vert^2\leq
g(\alpha)\,(1+4\,\alpha\,C_{6}(\al))\,\Vert p\psi_{\lambda}\Vert^2
+\alpha^2\,g(\alpha)\,C_{5}\;.
\end{equation}
 % ------------- %
Using our assumptions about $V$ we can write
 % ------------- %
\[(\psi_{\lambda};|V|\,\psi_{\lambda})=
-(\psi_{\lambda};V\,\psi_{\lambda})=\big(1\!-\!g(\alpha)\big)\,\Vert
p\psi_{\lambda}\Vert^2+|e_{\lambda}|\;,
\]
 % ------------- %
which yields an estimate to the last term at the right-hand side
of \eqref{eq:mu},
 % ------------- %
\begin{eqnarray}
\lefteqn{\Vert L^{-1/2}\Da\,p\, \psi_\lambda\Vert^2}\nonumber\\
&\geq&\Big[C_4(\mu)+\mu\,C_3(\mu)
-\frac{\lambda}{2}\,C_3(\mu)\,C\,\big(1\!-\!g(\alpha)\big)\,\Big]\,\Vert
p\,\psi_\lambda\Vert^2
-\frac{\lambda}{2}\,C_3(\mu)\,C\,|e_{\lambda}|\;;\nonumber\\
\label{fait2bis}
\end{eqnarray}
 % ------------- %
in case of (iii) this follows from \eqref{eq:hypV}, whereas for
(iii') we employ Lemma~\ref{lem:fait99}. Next we insert into
\eqref{fait1} from \eqref{evalP} and \eqref{fait2bis}; in
combination with Lemma~\ref{lem:1} we obtain
 % ------------- %
\begin{subequations}\label{energyter}
\begin{eqnarray}
\lefteqn{(\Psi;H_{\alpha}^V\,\Psi)}\nonumber\\
&\leq&\frac{\alpha}\pi\,\Vert\Psi\Vert^2
-\al^2\,\lao D\,D \A_{\alpha}^{-1}  \Da\, \Da\ora\label{eq:enerself}\\
&&-|e_\lambda|\,\Vert\Psi\Vert^2
+\alpha \,|e_\lambda|\,2\lambda\,C_3(\mu)\,C\,(1-a)\,\label{energy3-3}\\
&&+\left[ g(\alpha)\,\Big(1+4\,\alpha\,C_{6}(\alpha)
-2\,\alpha\,(1-a)\,\lambda\,C_3(\mu)\,C\Big)\right.\label{coeffg}\\
&& \quad\quad\left.-4\,\alpha \,(1-a)\,\big[C_4(\mu)+\mu\,C_3(\mu)
-\frac{\lambda}{2}\,C_3(\mu)\,C\,\big]\right]  \Vert p\,
\psi_\lambda\Vert^2\nonumber\\
&&+\alpha^2\,g(\alpha)\,C_{5}
-\alpha^2\,g(\al)\,C_1+\label{gal2}\\
&&+\frac{4\,\al^3}{a}\,\Vert L^{-1/2}P\,DL^{-1}\Da\,\Da \psi_\lambda\Vert^2\;
.\label{energy3-4}
\end{eqnarray}
\end{subequations}
 % ------------- %
Notice first that the term \eqref{energy3-4} behaves as
$\Ow(\al^3)$ for $\al\to 0$ which follows, e.g., from \cite[Lemma
15 (v)]{HHS}; thus it is irrelevant for the argument in the same
way as the shift coming from the infrared regularization.

The main idea is now to choose the function $g(\alpha)$ in such a
way that it cancels the factor in front of $\Vert p\,
\psi_\lambda\Vert^2$ in \eqref{coeffg}; this yields
 % ------------- %
\begin{equation}\label{eq:defg}
g(\alpha)=
\frac{4\,\alpha\,(1-a)\big[C_4(\mu)
+\mu\,C_3(\mu)-\frac{\lambda}{2}\,C_3(\mu)\,C\,\big]}
{1+4\,\alpha\,C_{6}(\alpha)
-2\,\alpha\,(1-a)\,\lambda\,C_3(\mu)\,C}\;.
\end{equation}
 % ------------- %
We choose also $\mu = \frac{\lambda}{2} C$ and fix the parameter
in \eqref{term1} by setting
 % ------------- %
$$1-a := \min\left\{\big(4C_4(\mu)\big)^{-1}, C_6(\al)\big(\mu
C_3(\mu)\big)^{-1}, 3/4\right\}\,; $$
 % ------------- %
this yields $g(\al)\leq \al$ which means that  \eqref{gal2} $ =
\Ow(\alpha^3)$.

On the other hand, since
 % ------------- %
\begin{eqnarray}
\Vert \Psi\Vert^2 &\!=\!& 1+4\,\alpha\,\Vert
L^{-1}\Da\,p\psi_{\lambda}\Vert^2+\alpha^2\,\Vert
L^{-1}\Da\,\Da\,\psi_{\lambda}\Vert^2 \nonumber \\
&\!=\!& 1+\Ow(\alpha)\;,
\end{eqnarray}
 % ------------- %
we deduce from \eqref{eq:self} that
 % ------------- %
\[
\eqref{eq:enerself}=E(\alpha, 0)\,\Vert\Psi\Vert^2+\Ow(\alpha^3)\;.
\]
 % ------------- %
We denote by $E(\beta)$ the bottom of the spectrum of
$p^2+\beta\,V$, \emph{i.e.}
 % ------------- %
$$ E(\beta) := \infspec (p^2+\beta\,V).$$
 % ------------- %
We have $E(\lambda_{0})=0$ by assumption, and since the ground
state represents always case (A) in the terminology of \cite{KS},
in other words, zero is not an eigenvalue of $p^2+\lambda_{0}\,V$,
we know that
 % ------------- %
\begin{equation}\label{aV}
E(\beta)= -b(V)\, (\beta-\lambda_{0})^2+\Ow\big(
(\beta-\lambda_{0})^3\big)\;,
\end{equation}
 % ------------- %
holds for $\beta\geq \lambda_{0} $, close to $\lambda_{0}$, and
for some positive constant $b(V)$ depending only on the potential
$V$. Notice that the above asymptotic expansion coming from
\cite[Theorem~2.3]{KS} was derived there for $V\in
C_0^\infty(\R^3)$, however, an extension to the Rollnik class is
straightforward. Recall now that
 % ------------- %
\[|e_{\lambda}|=-\big(1\!-\!g(\alpha)\big)\,E\big(\lambda\,
(1\!-\!g(\alpha))^{-1}\big)\;.\]
 % ------------- %
Since $(1\!-\!g(\alpha))\lambda_{0}<\lambda\le \lambda_{0}$ holds
by assumption and $g(\alpha)=\Ow(\alpha)$, we have
 % ------------- %
\[
\lambda\,\big(1\!-\!g(\alpha)\big)^{-1}-\lambda_{0}\leq
\lambda_{0}
g(\alpha)\,\big(1\!-\!g(\alpha)\big)^{-1}=\Ow(\alpha)\,,
\]
 % ------------- %
and therefore
 % ------------- %
\[
|e_{\lambda}|= b(V) \left(\frac{\lambda}{1\!-\!g(\alpha)} -
\lambda_{0} \right)^2 + \Ow(\alpha^3)\,,
\]
 % ------------- %
where the first term at the right-hand side is $\Ow(\alpha^2)$.
Returning to \eqref{energyter} we conclude from the last claim
that the second term in \eqref{energy3-3} is of order of
$\alpha^3$. This yields for all $\lambda$ in the considered range
and for small $\alpha$ the asymptotic inequality
 % ------------- %
\[
E(\alpha,\lambda\,V)\leq E(\alpha, 0)
-|e_{\lambda}|+\Ow(\alpha^3)\;.\]
 % ------------- %
Since $b(V)>0$ the second term at the right-hand side is negative
and dominates over the error for $\alpha$ sufficiently small. This
demonstrates that the sought inequality \eqref{eq:crit} is valid
under the assumptions we have made and thus it proves
Theorem~\ref{mth}. \vskip6pt

%%%%%%%%%%%%%%%%%%%%%%%

\end{document}